\documentclass[%
reprint,
amsmath,amssymb,
aps,
superscriptaddress,
prc
]{revtex4-1}

\usepackage{graphicx}
\usepackage{dcolumn}
\usepackage{bm}
\usepackage{hyperref}

\usepackage[capitalise]{cleveref}
\usepackage{comment}
\usepackage{color}

\begin{document}
	
	
	\title{Extracting the kinetic freeze-out properties of high energy pp collisions at the LHC with event shape classifiers}
	
	\author{Jialin He}
	\affiliation{School of Mathematics and Physics, China University of
		Geosciences (Wuhan), Wuhan 430074, China}
	\author{Xinye Peng}
	\affiliation{School of Mathematics and Physics, China University of
		Geosciences (Wuhan), Wuhan 430074, China}
	\affiliation{Key Laboratory of Quark and Lepton Physics (MOE) and Institute
		of Particle Physics, Central China Normal University, Wuhan 430079, China}
        \author{Zhongbao Yin}
	\affiliation{Key Laboratory of Quark and Lepton Physics (MOE) and Institute
		of Particle Physics, Central China Normal University, Wuhan 430079, China}        
	\author{Liang Zheng}\email{zhengliang@cug.edu.cn}
	\affiliation{School of Mathematics and Physics, China University of
		Geosciences (Wuhan), Wuhan 430074, China}
	\affiliation{Shanghai Research Center for Theoretical Nuclear Physics, NSFC and Fudan University, Shanghai 200438, China}
	\affiliation{Key Laboratory of Quark and Lepton Physics (MOE) and Institute
	of Particle Physics, Central China Normal University, Wuhan 430079, China}	
	
	\date{\today}
	
	\begin{abstract}

    Event shape measurements are crucial for understanding the underlying event and multiple-parton interactions (MPIs) in high energy proton-proton (pp) collisions. In this paper, the Tsallis Blast-Wave model with independent non-extensive parameters for mesons and baryons, was applied to analyze transverse momentum spectra of charged pions, kaons, and protons in pp collision events at $\sqrt{s}=13$ TeV classified by event shape estimators relative transverse event activity, unweighted transverse spherocity, and flattenicity. Our analysis reveals consistent trends in the kinetic freeze-out temperature and non-extensive parameter across different collision systems and event shape classes.
    The use of diverse event-shape observables in pp collisions has significantly expanded the accessible freeze-out parameter space, allowing for a more comprehensive exploration of its boundaries. Among these event shape classifiers, flattenicity emerges as a unique observable for disentangling hard process contributions from additive MPI effects, allowing the isolation of collective motion effects encoded by the radial flow velocity.
    Through the analysis of the interplay between event-shape measurements and kinetic freeze-out properties, we gain deeper insights into the mechanisms responsible for flow-like signatures in pp collisions.

	\end{abstract}
	

	\maketitle
	
	
	\section{Introduction}
	\label{sec:level1}
	
Lattice Quantum Chromodynamics (QCD) calculations indicate that regions of extremely high energy density are likely to be created in high energy heavy ion collisions~\cite{Broniowski:2008vp,Elfner:2022iae}. In these regions, quarks and gluons confined in nucleons can be released, forming a nearly perfect quark-gluon plasma (QGP) with high temperature and small viscosity~\cite{Harris:2023tti}. The thermalization of the deconfined nuclear matter occurs on an extremely short timescale and rapidly expands via the hydrodynamic evolution process~\cite{Heinz:2009xj,Gale:2013da}. During the expansion of the QGP medium, the temperature of the system drops to a level where the quark and gluon degrees of freedom begin to freeze and form final-state hadrons~\cite{Retiere:2003kf}. The significant collective motion of the QGP matter during the hydrodynamic evolution may lead to sizable collectivity effects in experiments embedded in the final state hadron momentum distributions. Analyzing the experimentally observed collective flow behavior of final-state hadrons provides valuable information about the evolution and phase transitions of QGP matter~\cite{Schnedermann:1993ws,Schnedermann:1994gc,Heinz:2004qz,Chen:2024aom,Shou:2024uga}.

Recent observations of collective flow-like behaviors in small collision systems, such as high-multiplicity proton-proton (pp) events, have generated considerable discussion regarding the potential formation of QGP matter in these collisions where genuine fluid-like dynamics are unexpected~\cite{Nagle:2018nvi,Adolfsson:2020dhm,Noronha:2024dtq}. It is speculated that hot spots arising from high initial energy density in small systems may give rise to sizable collective flow and strong temperature fluctuations. In pp collisions, the formation of these hot spots is believed to be linked to multiple-parton interactions (MPIs), defined as the occurrence of multiple independent partonic scatterings within a single hadronic collision~\cite{Sjostrand:2004pf}. This process contributes significantly to the underlying event by enhancing particle multiplicity and energy deposition in localized regions. The cumulative effects of MPI can create conditions resembling those seen in larger collision systems~\cite{dEnterria:2010xip}. High MPI activity leads to denser particle environments and the isotropization of particle distributions due to the superposition of multiple scatterings which can amplify final-state interactions or mimic azimuthal anisotropies traditionally associated with hydrodynamic flow. 

Experimentally constraining MPI effects is vital to disentangle their contributions from genuine collective flow and other non-flow correlations in the final state. The number of charged particles within certain detector acceptance region were initially proposed to study the MPI dependent effects and used to compare the flow signals across different collision systems. However, this observable is found to bias the high multiplicity jets in the corresponding detector region~\cite{ALICE:2022qxg,ALICE:2018pal}. Measurements of event shape observables for small systems, such as transverse event activity~\cite{Martin:2016igp,ALICE:2023yuk}, spherocity~\cite{Banfi:2010xy,ALICE:2012cor}, and charged particle flattenicity~\cite{ALICE:2019dfi,Ortiz:2022mfv}, have been proposed to identify events with less bias to jet productions but more sensitive to the MPI related contributions. By correlating these event shape observables with flow-sensitive measurements, such as the elliptic flow coefficient and long-range correlations, experimental collaborations have begun to quantify the extent to which MPI shape the observed signals~\cite{Ortiz:2017jho,ALICE:2023bga,ALICE:2024apz}.

In this work, we extract the kinetic freeze-out temperature and the radial flow velocity of collision systems via applying the Tsallis Blast-Wave (TBW) analysis~\cite{Tang:2008ud,Liu:2022ikt} to the transverse momentum spectra in high energy pp collision events with different event shape classifiers. The imprints of the initial fluctuations and the viscosity of the expanding medium can be embedded in the non-extensive parameter in the TBW model and its correlation with temperature and flow velocity~\cite{Wilk:1999dr,Wilk:2008ue,Chen:2020zuw,Che:2020fbz,Gu:2022xjn}. 
By systematically comparing the freeze-out parameters extracted using various event shape classifiers, we directly assess their effectiveness in event shape engineering and their sensitivity to isolate underlying collective dynamics for small system. It is expected that the non-extensive parameter, which characterizes deviations from local thermal equilibrium, will decrease in more isotropic and MPI dominated events, reflecting a system closer to thermalization. In contrast, jet dominated events are expected to exhibit enhanced non-equilibrium features and reduced collective flow signatures. Through this comparative analysis, we aim to determine the extent to which each event shape observable can disentangle soft, flow like behavior from contributions associated with hard scattering processes.
Comparing the extracted parameters from different event shape variables is important to understand the sensitivity of these event shape control variables to the underlying collective dynamics in pp collisions~\cite{Prasad:2024gqq}. Exploring the kinetic freeze-out features varying with the event shapes can be of great interest to quantify the MPI effect in generating the flow like effects in small systems and shed light on the understanding of the origin of the collectivity like behavior observed in small systems.

    The rest of this paper is organized as follows: Sect.\ref{sec:formalism} briefly describes the event shape classifiers used in this analysis and the key parameters in our TBW fit framework. Sect.~\ref{sec:results} compares the extracted freeze-out properties with different event shapes, and Sect.~\ref{sec:summary} summarizes the main conclusions.
  
	\section{\label{sec:formalism}Research approach  }
    \subsection{Relative transverse event activity}

    Event activity measurements are indispensable for probing the underlying event properties and understanding the MPI effects in high-energy proton-proton collisions. It is usually studied by analyzing the particle production in azimuthal regions relative to the leading particle direction in one event. With the trigger particle being the one with the largest $p_T$ in an event and the rest termed as associated particles, the azimuthal plane can be divided into three different topological regions, defined by the angular difference (\( \mid\Delta\phi\mid= |\phi_{\text{trig}} - \phi_{\text{assoc}}|  \)) between the trigger and the associate particle~\cite{CDF:2001onq, ALICE:2023yuk}. The toward region \( |\Delta\phi| < 60^\circ \) is expected to be associated with the main jet production in an event. The away region \( |\Delta\phi| \geq 120^\circ \) contains particles fragmented from the recoil jet. The particles in the transverse region \( 60^\circ \leq |\Delta\phi| < 120^\circ \) predominantly come from the underlying event process, which are subject to various sources beyond jet fragmentation, including initial- and final-state radiation, beam remnants, and MPIs~\cite{Sjostrand:1987su}. Therefore, the relative transverse activity (\( R_T \))~\cite{Martin:2016igp}, built from the transverse region which is expected to have low sensitivity to the hard processes, has been proposed to classify events and gain insight into the modifications to the charged hadron $p_T$ spectra
    \begin{eqnarray}\label{eq:rt}
        R_T = N_T / \langle N_T \rangle ,
    \end{eqnarray}
    where \( N_T \) is the number of charged particles measured in the transverse region for each event, and \( \langle N_T \rangle \) is the average number of charged particles across all analyzed events. The experimental data used in this study divides \( R_T \) into four intervals: 0–0.5, 0.5–1.5, 1.5–2.5, and 2.5–5.~\cite{ALICE:2023yuk}. Events are selected by requiring a leading charged particle with $p_T>5$ GeV$/c$ at mid-rapidity. The transverse momentum spectra of identified hadrons are analyzed separately in the toward, away, and transverse regions, defined relative to the azimuthal angle of the leading particle. Increasing \( R_T \) represents a transition to the multiparton interaction dominated underlying event categories.

    \subsection{Unweighted transverse spherocity}
In high-energy hadronic collisions, spherocity is a key event-shape observable used to characterize the geometrical shapes of the particle distributions in the transverse plane~\cite{Banfi:2010xy}. It distinguishes between jet-like events, featuring collimated, high-momentum particle productions, and isotropic events, where particles are uniformly distributed signaling softer processes. Unlike the sphericity measurement\cite{ALICE:2012cor}, which assesses three-dimensional isotropy and is more suitable for $e^+e^-$ collisions, spherocity is well-suited to hadron colliders, in which the transverse momentum information is of great interest. In the traditional spherocity calculation~\cite{ALICE:2019dfi}, tracks with high \( p_T \) contribute disproportionately, leading to significant differences between neutral and charged particles in jet-like events. 
This study focuses on the unweighted transverse spherocity, where all particles contribute equally, regardless of their transverse momentum. By suppressing jet contributions, unweighted spherocity enhances sensitivity to soft and bulk particle production.
Such an approach is particularly effective in small systems, where separating jet-induced events from those potentially exhibiting collective behavior is crucial. We employ the $p_T$ data provided in 
Ref.\cite{ALICE:2023bga}, categorized by the unweighted transverse spherocity estimator ($S_{O}^{p_{T}=1}$). It is calculated as
    \begin{eqnarray}\label{eq:sopt}
        S_{O}^{p_{T}=1} = \frac{\pi^2}{4} \min_{\hat{n}} \left( \frac{\sum_i |\hat{p}_{T,i} \times \hat{n}|}{N_{\text{trks}}} \right)^2,
    \end{eqnarray}
    where the summation is over all charged particles with transverse momentum \( p_T > 0.15 \, \text{GeV/c} \), \( \hat{p}_T \) represents the unit vector of transverse momentum, \( N_{\text{trks}} \) is the number of charged particles in a given event, and \( \hat{n} \) is the unit vector that minimizes \( S_{O}^{p_{T}=1} \). This \( S_{O}^{p_{T}=1} \) definition treats all charged tracks with equal weight, setting \( p_T = 1 \) for each, in contrast to the original transvserse momentum weighted spherocity formulation.
    To constrain the hardness of the events more effectively, the unweighted spherocity data~\cite{ALICE:2023bga} used in this work is obtained using mid-rapidity multiplicity estimator in tandem with $S_{O}^{p_{T}=1}$ selection divided into three categories: 0–10\%, 90–100\%, and 0–100\% (\(S_{O}^{p_{T}=1} \)-integrated). The 0–10\% interval indicates low spherocity values corresponding to jet-like events with particles predominantly aligned in the azimuthal plane, and the 90–100\% interval denotes high spherocity values associated with isotropic events where particles are uniformly distributed in the azimuthal plane. Only high multiplicity events in the top 1\% mid-rapidity multiplicity percentile are used in this data.  

    \subsection{Flattenicity}
Flattenicity is a new event-shape observable developed by the ALICE Collaboration to measure local charged-particle multiplicity fluctuations in the forward V0 detector on an event-by-event basis~\cite{ALICE:2019dfi}.
Unlike traditional multiplicity estimators, which focus on total yield and can be biased by multi-jet events, flattenicity targets density variations without momentum weighting, reducing bias from high $p_T$ jets and enhancing sensitivity to soft particle production~\cite{Ortiz:2022mfv}. The flattenicity measurement obtained by ALICE is performed using the forward V0 scintillator arrays~\cite{ALICE:2013axi}, in which the distribution of charged-particle multiplicities across the pseudorapidity ($\eta$) and azimuthal angle ($\phi$) phase space are segmented into $N_{\text{cell}} = 64$ elementary cells.  . The flattenicity ($\rho$) is defined as follows:
    \begin{eqnarray}\label{eq:rho}
        \rho = \frac{\sqrt{\sum_{i=1}^{64} \left( N_{\text{ch}}^{\text{cell},i} - \langle N_{\text{ch}}^{\text{cell}} \rangle \right)^2 / N_{\text{cell}}^2}}{\langle N_{\text{ch}}^{\text{cell}} \rangle},
    \end{eqnarray}
    where $N_{\text{ch}}^{\text{cell},i}$ represents the particle multiplicity in the $i^{\text{th}}$ cell, and $\langle N_{\text{ch}}^{\text{cell}} \rangle$ is the average multiplicity across all 64 cells in each event. This formula quantifies how evenly particles are distributed across the $\eta$-$\phi$ cells. A small $\rho$ value indicates a uniform multiplicity distribution, suggesting isotropic particle production, while a large $\rho$ value points to significant fluctuations, possibly due to clustered activity such as jets or hard processes. 
    Comparisons with PYTHIA Monte Carlo models indicate that flattenicity correlates with the number of multiparton interactions and behaves differently than traditional V0M multiplicity estimators, making it a powerful tool to disentangle MPI-driven fluctuations from collective phenomena.    
    In experiments, the results are commonly expressed in terms of $1 - \rho$ so that higher values correspond to more isotropic events, thus aligning flattenicity’s orientation with the conventions of other event-shape observables.
    The distribution of $1 - \rho$ is then divided into several percentile intervals with the lowest percentile (e.g. 0-1\%) capturing the flattest events with maximal MPIs, whereas the highest percentile (e.g. 50-100\%) selects the bumpy events with few MPIs, thus providing a uniform framework that mitigates biases from hard scattering when studying soft QCD dynamics.

    Our current study employs the data from Ref.~\cite{ALICE:2024vaf}, in which a double-differential event classification based on both multiplicity and flattenicity is applied. We include the $p_T$ spectra from both the minimum bias sample, encompassing all inelastic collisions (V0M percentile 0–100\%), and from the high multiplicity bin, consisting of the top 1\% of events with the highest multiplicity (V0M percentile 0–1\%). Within each category, events are further classified according to their flattenicity values. This approach enables us to explore the multiplicity-dependent nature of flattenicity across a broad spectrum of event types, offering insights into the dynamics of high-energy collisions.

    \subsection{Tsallis Blast-Wave model}

    In the Tsallis Blast‐Wave (TBW) model, the invariant differential yield is obtained by integrating a Tsallis‐distributed source over longitudinal rapidity, azimuthal angle, and transverse radius, thus combining collective expansion with non‐extensive statistics to describe the full $p_T$ spectrum in high‐energy collisions.
    The invariant differential particle yield of a hadron with mass \( m \) within TBW framework can be expressed as:
    \begin{eqnarray}\label{eq:TBW}
	\frac{d^{2}N}{2\pi m_{T}dm_{T}dy}|_{y=0} & = & A\int^{+y_{b}}_{-y_{b}}m_{T}\cosh(y_{s})dy_{s}\int^{\pi}_{-\pi}d\phi \\ \nonumber& \times & \int^{R}_{0}rdr[1+\frac{q-1}{T}(m_T\cosh(y_s)\cosh(\rho) \\ \nonumber
	& &-p_T\sinh(\rho)\cos(\phi))]^{-1/(q-1)}.
	\end{eqnarray}
    In the above equation, \( T \) represents the freeze-out temperature of the expanding source. \( R \) is the boundary along the transverse radial direction (edge of the hard sphere). \( A \) is a normalization constant. \( m_{T} = \sqrt{p^{2}_{T} + m^{2}} \) denotes the transverse mass of the particle. \( y_{s} \) represents the rapidity of the source, \( y_{b} \) represents the beam rapidity. \( \phi \) is the angle of particle emission relative to the fluid flow velocity. \( \rho = \tanh^{-1} \beta(r) \) defines the radial flow profile, described by the transverse flow \( \beta(r) = \beta_{S} \left( \frac{r}{R} \right)^n \), where \( n \) is the flow profile index. \( \langle \beta \rangle = \beta_{S} \cdot \frac{2}{2+n} \) represents the average transverse flow velocity. Setting \( n = 1 \) yields a linear velocity profile, resulting in \( \langle \beta \rangle = \frac{2}{3} \beta_{S} \)~\cite{Liu:2022ikt}. 
    To account for blue‐shift effects due to collective expansion, an effective temperature \( T_{eff} = \sqrt{\frac{1 + \langle \beta \rangle}{1 - \langle \beta \rangle}} T \) is often introduced, reflecting the observed $p_T$ spectra hardening. 
    In the TBW4 variant, the TBW model is extended by introducing independent non‐extensivity parameters $q_M$ for mesons and $q_B$ for baryons~\cite{Tang:2008ud,Jiang:2013gxa}. This modification yields a markedly improved fit to identified hadron $p_T$ spectra in small collision systems~\cite{Che:2020fbz,Liu:2022ikt}. This performance enhancement underscores the pivotal role of baryon‐number–dependent non‐equilibrium dynamics in small systems, where fragmentation and collective effects interplay in unique ways. In this work, we perform the Blast-Wave analysis of the $p_T$ spectra of pions, kaons and protons at mid-rapidity in pp collisions with different event shapes using the TBW4 fit. Restricting the analysis to these particles allows for a more controlled and meaningful extraction of the bulk freeze-out dynamics. In the end, we would like to emphasize that the use of the Tsallis Blast-Wave model in this work provides a flexible yet phenomenological framework to characterize freeze-out properties in small systems. While the TBW model captures key features of the transverse momentum spectra, it lacks a microscopic foundation and assumes radial symmetry in both the velocity and temperature fields of the expanding source, which might be a simplification especially for the strongly fluctuating asymmetric dynamics present in individual pp collisions dominated by jets. The extracted parameters should therefore be interpreted as effective quantities that reflect the combined influence of thermal motion, radial flow, and non-equilibrium fluctuations. These limitations will motivate future work incorporating more differential modeling with microscopic evolution mechanism and experimental correlation analyses~\cite{Zhang:2025pqu}.

	\begin{figure*}[htbp]
		\begin{center}
			\includegraphics[width=1.0\textwidth]{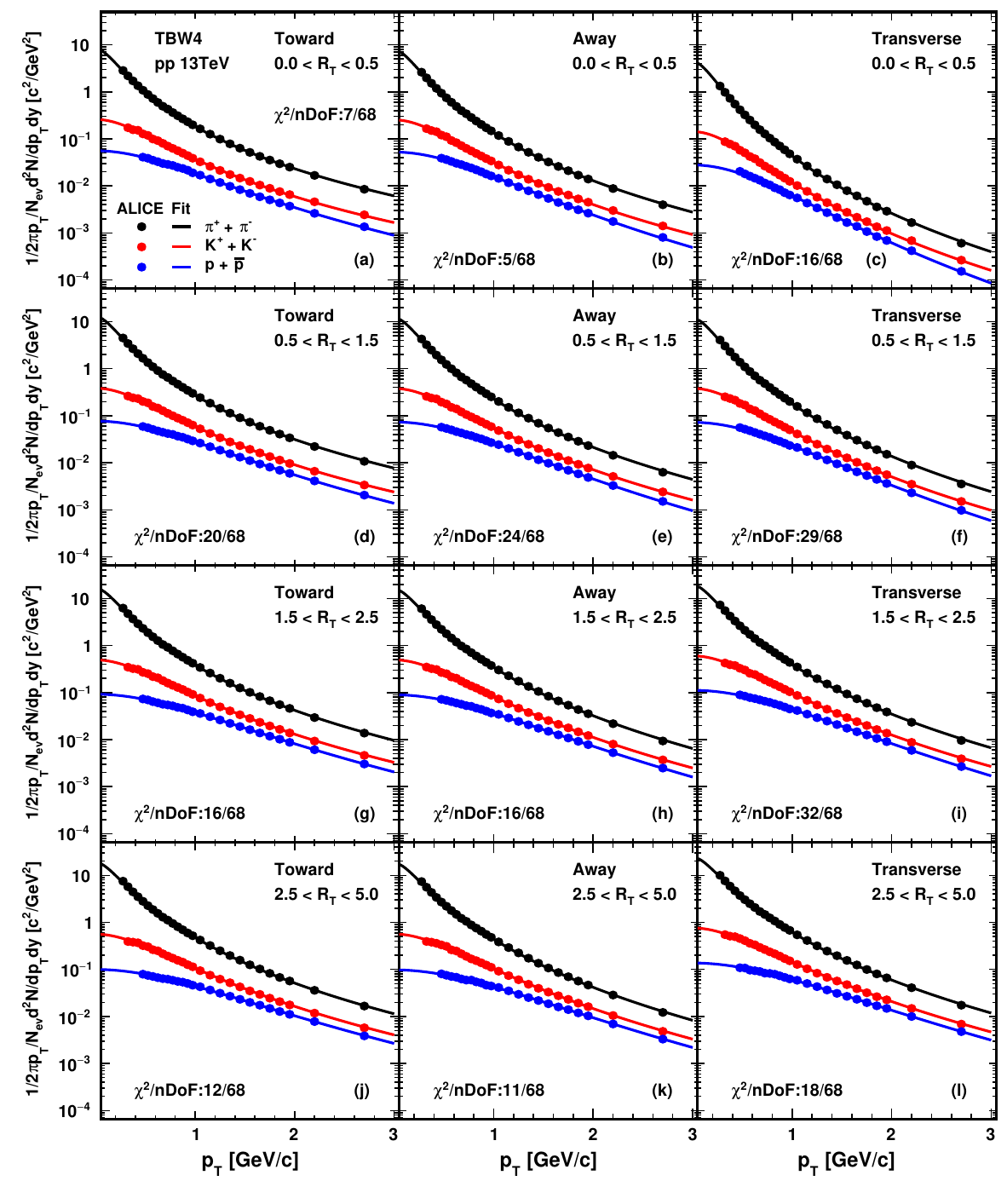}
			     \caption{The TBW4 fits to hadron spectra in pp collisions at $\sqrt{s}=$ 13~TeV. Black, red, and blue correspond to $\pi$, $K$, and $p$ particles, respectively. The points represent the ALICE experimental data~\cite{ALICE:2023yuk}, and the lines represent the fit results. From top to bottom, the rows represent the $R_T$ intervals of 0-0.5, 0.5-1.5, 1.5-2.5, and 2.5-5. From left to right, the columns represent the forward, backward, and transverse regions. The uncertainties in the experimental data are the quadratic sum of statistical and systematic errors.}
            \label{fig:pt_region}
		\end{center}
	\end{figure*}
	
	\begin{figure*}[htbp]
		\begin{center}
			\includegraphics[width=1.0\textwidth]{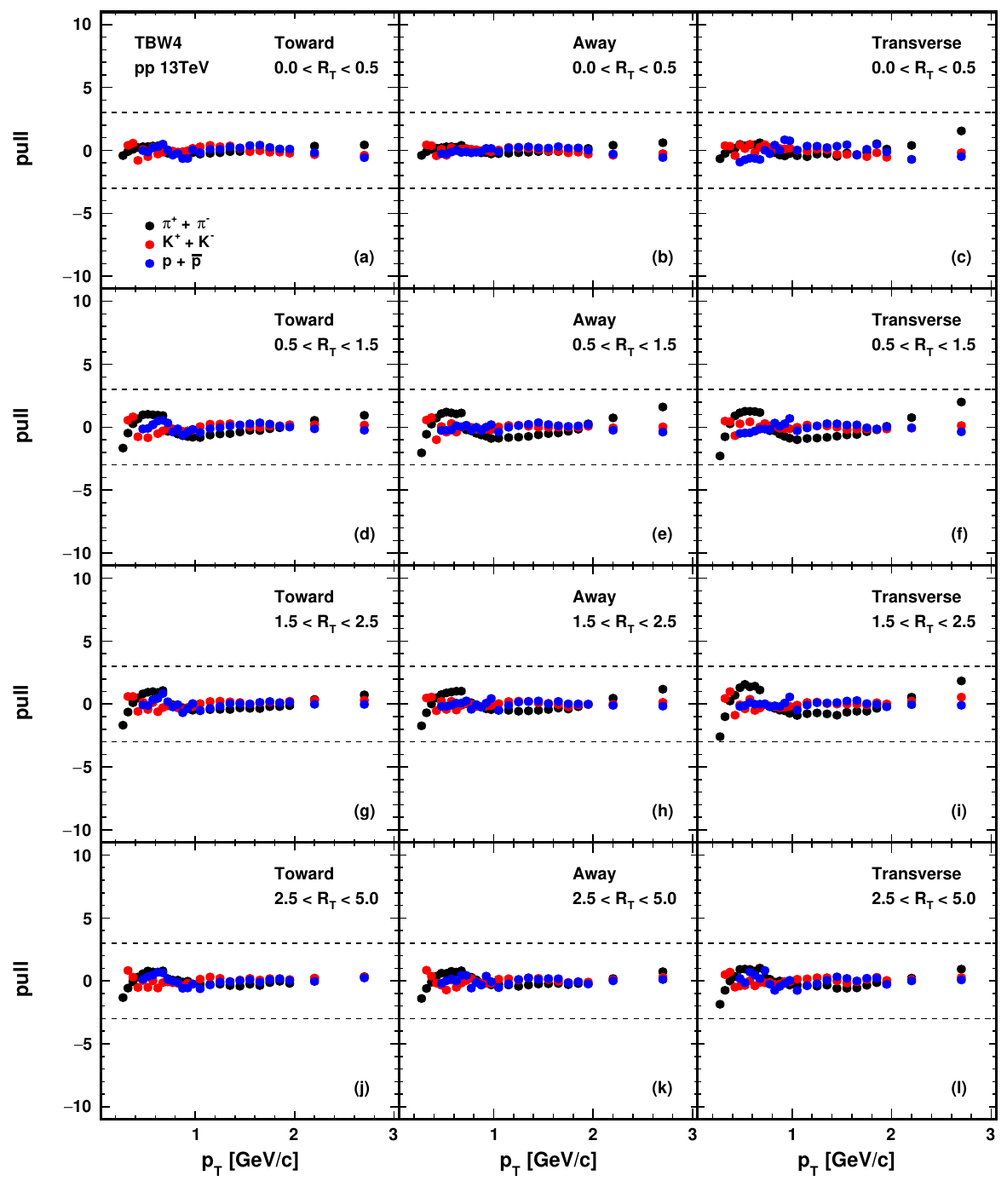}
			     \caption{The deviations of TBW4 fits to hadron spectra divided by data uncertainties in pp collisions at $\sqrt{s}=$ 13~TeV. Black, red, and blue correspond to $\pi$, $K$, and $p$ particles, respectively. From top to bottom, the rows represent the $R_T$ intervals of 0-0.5, 0.5-1.5, 1.5-2.5, and 2.5-5. From left to right, the columns represent the forward, backward, and transverse regions. The dashed lines represent where the difference between model and experiment data is three times the error of data.
            }\label{fig:pull_region}
		\end{center}
	\end{figure*}

    \begin{figure*}[htbp]
		\begin{center}
			\includegraphics[width=1.0\textwidth]{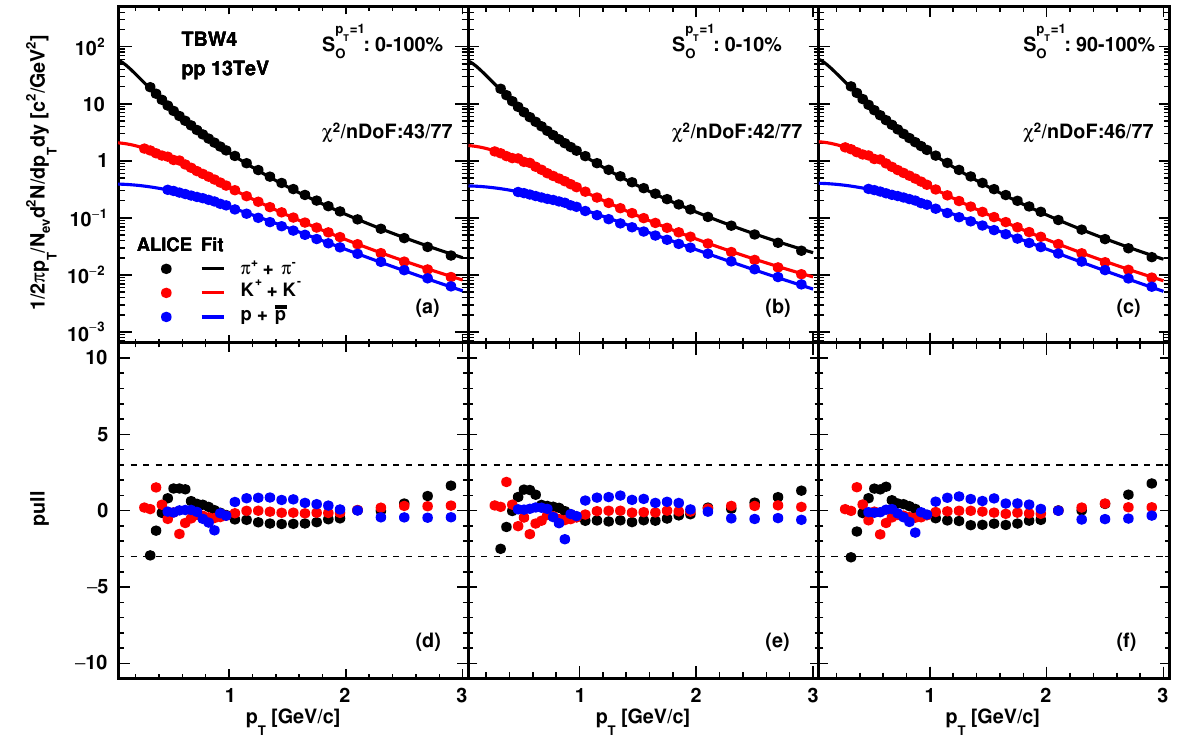}
			\caption{The TBW4 fits to hadron spectra in pp collisions at $\sqrt{s}=$ 13 TeV. Black, red, and blue correspond to $\pi$, $K$, and $p$ particles, respectively. The points represent the ALICE experimental data~\cite{ALICE:2023bga}, and the lines represent the fit results.From left to right, the columns represent sphericity intervals of 0-100\%, 0-10\%, and 90-100\%. The top row shows the $p_T$ spectra, and the uncertainties in the experimental data are the quadratic sum of statistical and systematic errors. The Bottom show the deviations from the $p_T$ fit, and the dashed lines indicate where the difference between the model and experimental data is three times the data's uncertainty.}\label{fig:sopt}
		\end{center}
	\end{figure*}

    \begin{figure*}[htbp]
		\begin{center}
			\includegraphics[width=1.0\textwidth]{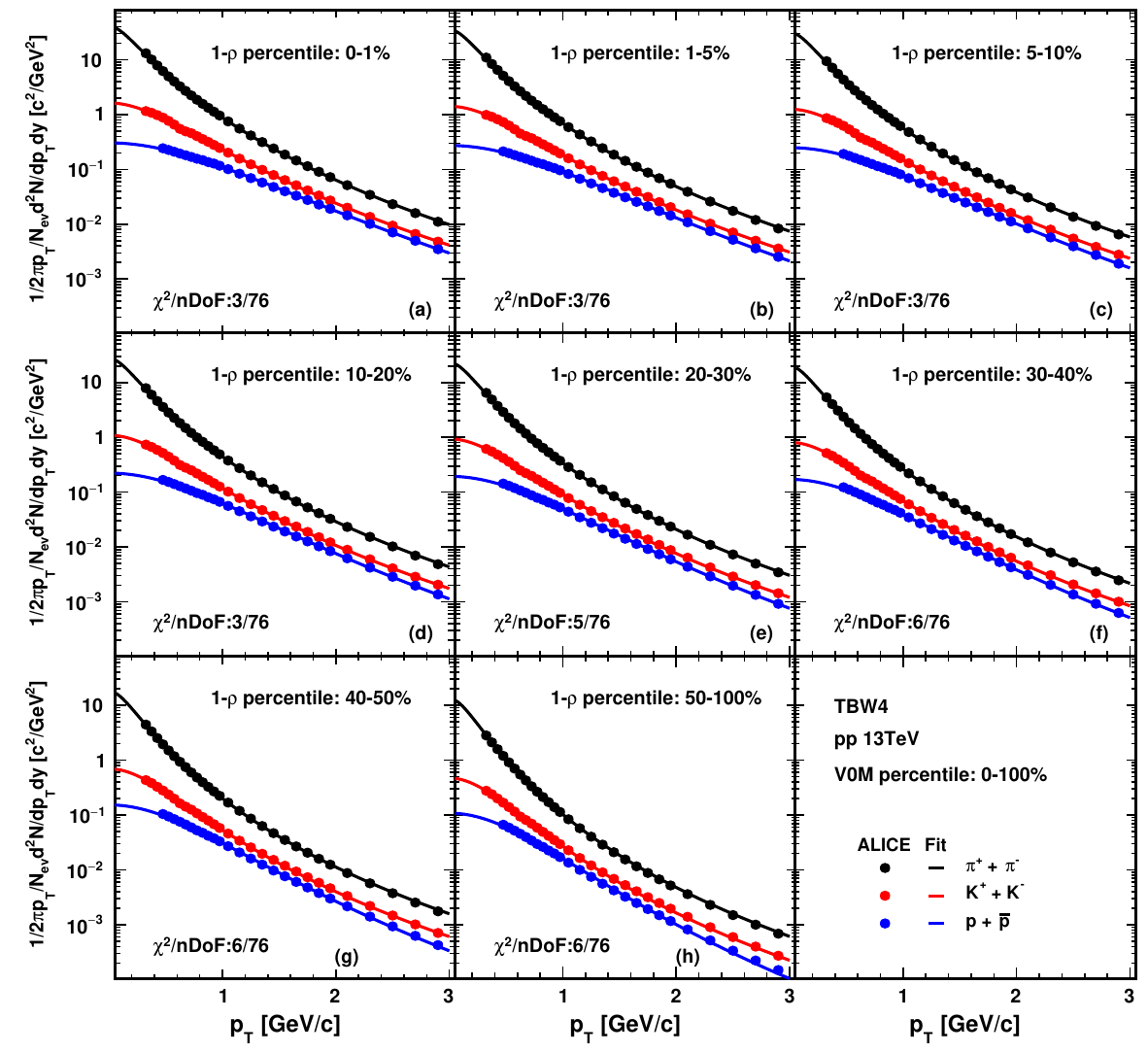}
			\caption{The TBW4 fits to hadron spectra in pp collisions at $\sqrt{s}=$ 13~TeV. Black, red, and blue correspond to $\pi$, $K$, and $p$ particles, respectively. The points represent the ALICE experimental data~\cite{ALICE:2024vaf}, and the lines represent the fit results. From panel (a) to (h), the results of different flattenicity event classes for multiplicity-integrated events (V0M percentile $0–100\%$) are presented. The uncertainties in the experimental data are the quadratic sum of statistical and systematic errors.}\label{fig:pt_rho}
		\end{center}
	\end{figure*}

    \begin{figure*}[htbp]
		\begin{center}
			\includegraphics[width=1.0\textwidth]{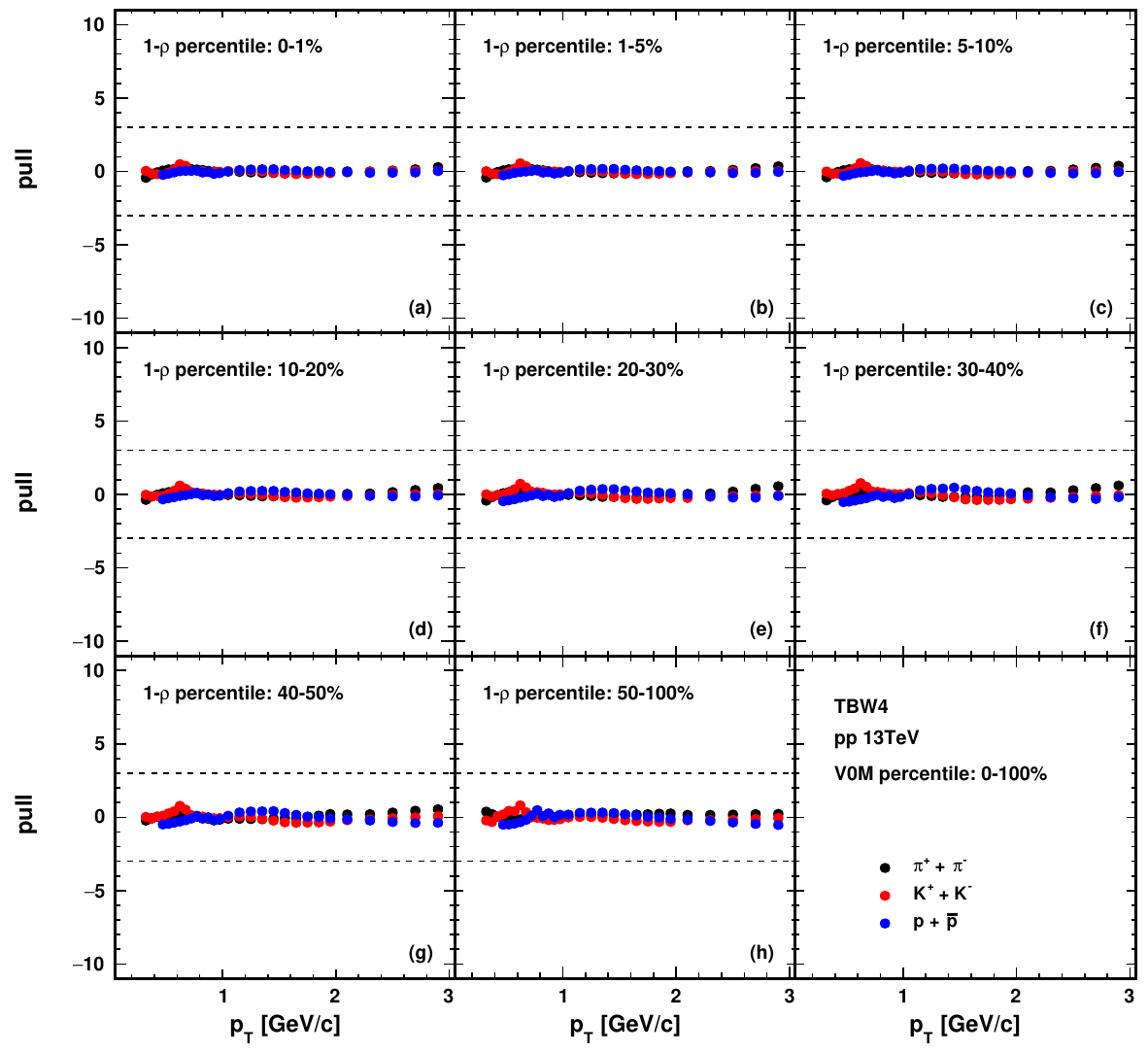}
			\caption{The deviations of TBW4 fits to hadron spectra divided by data uncertainties in pp collisions at $\sqrt{s}=$ 13~TeV. Black, red, and blue correspond to $\pi$, $K$, and $p$ particles, respectively. From From panel (a) to (h), the results of different flattenicity event classes for multiplicity-integrated events (V0M percentile $0–100\%$) are presented. The dashed lines represent where the difference between model and experiment data is three times the error of data.
            }\label{fig:pull_rho}
		\end{center}
	\end{figure*}

    \begin{figure*}[htbp]
		\begin{center}
			\includegraphics[width=1.0\textwidth]{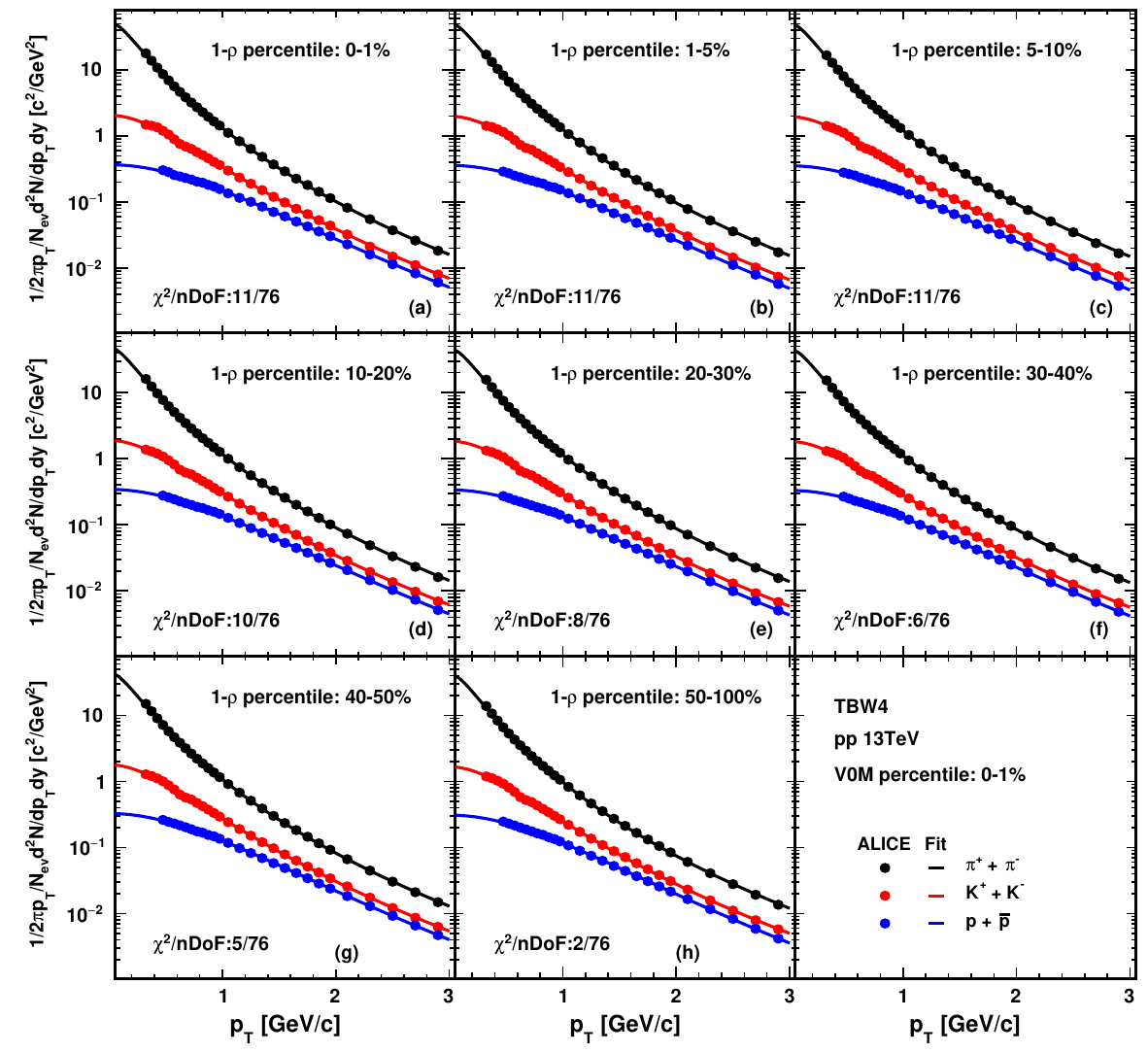}
			\caption{The TBW4 fits to hadron spectra in pp collisions at $\sqrt{s}=$ 13~TeV. Black, red, and blue correspond to $\pi$, $K$, and $p$ particles, respectively. The points represent the ALICE experimental data~\cite{ALICE:2024vaf}, and the lines represent the fit results. From panel (a) to (h), the results of different flattenicity event classes for high-multiplicity events (V0M percentile 0-1\%) are presented. The uncertainties in the experimental data are the quadratic sum of statistical and systematic errors.
            }\label{fig:pt_rhoHM}
		\end{center}
	\end{figure*}

    \begin{figure*}[htbp]
		\begin{center}
			\includegraphics[width=1.0\textwidth]{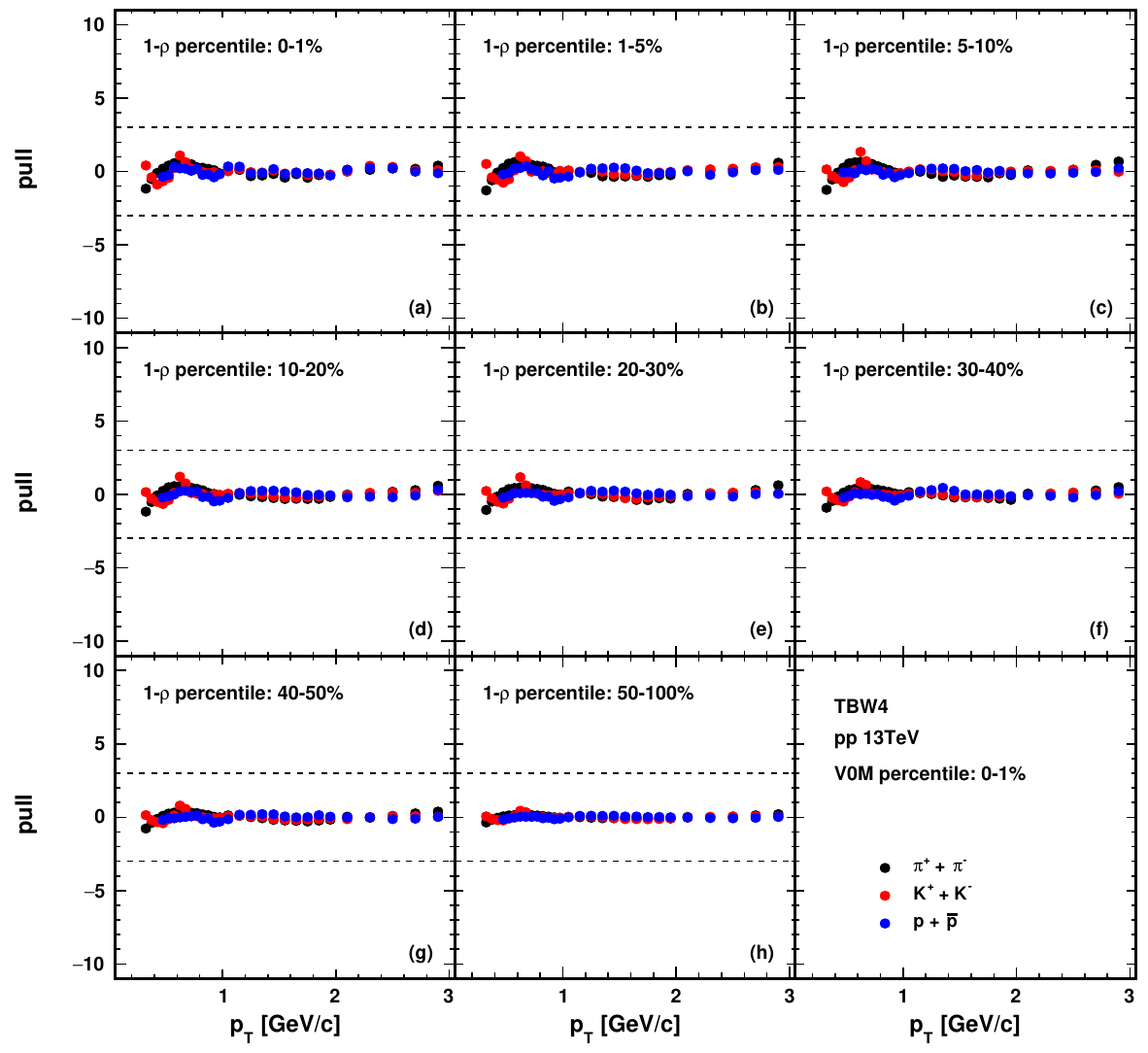}
			\caption{The deviations of TBW4 fits to hadron spectra divided by data uncertainties in pp collisions at $\sqrt{s}=$ 13~TeV. Black, red, and blue correspond to $\pi$, $K$, and $p$ particles respectively. From panel (a) to (h), the results of different flattenicity event classes for high-multiplicity events (V0M percentile $0–1\%$) are presented. The dashed lines represent where the difference between model and experiment data is three times the error of data.
            }\label{fig:pull_rhoHM}
		\end{center}
	\end{figure*}
    
    \section{\label{sec:results}Results}
	\subsection{Transverse momentum spectra}
This section compares the Tsallis Blast-Wave model fits considering independent baryon and meson non-extensive parameters of the transverse momentum spectra for charged pions, kaons and protons in pp collisions at $\sqrt{s}=$13 TeV using different event shape classifiers. To ensure consistent bulk property extraction, we only include transverse momentum data within $p_T<3$ $\mathrm{GeV}/c$ region in our analysis. The average flow velocity is constrained to $0<\langle\beta\rangle<2/3$ to eliminate the non-physical parameter regime. To quantify deviations between model fits and experimental data, we calculate the Pull distribution defined as $pull=(fit-data)/(data\ error)$. Positive (negative) values indicate where the fit overestimates (underestimates) the data. The Pull magnitude directly corresponds to the number of standard deviations between the fit and experimental measurements.

In Fig.~\ref{fig:pt_region}, we present the comparison of TBW4 fits to the $p_T$ spectra data~\cite{ALICE:2023yuk} divided into different event topology regions, constrained by transverse event activity $R_T$ in each region. Events used in this analysis are selected by requiring a leading charged particle with $p_T>5$ GeV$/c$ at mid-rapidity. The event topology regions are defined relative to the azimuthal angle of this leading particle.
The results are displayed in panels organized from left to right as the toward, away, and transverse regions. From top to bottom, the panels correspond to $R_T$ intervals of 0–0.5, 0.5–1.5, 1.5–2.5, and 2.5–5. Experimental data points and theoretical fits are represented by markers and curves respectively, where black, red, and blue colors correspond to pions, kaons, and protons.  The $\chi^2/nDOF$ values obtained from each fit are indicated within the panels, with further details of the fit parameters provided in Tab.~\ref{tabTBW4}.

    It is found in Fig.~\ref{fig:pt_region} that the features of the particle yield dependent on the event topology region and the $R_T$ variation are generally well reproduced in the TBW4 fits. Considering that these particle $p_T$ spectra are selected by requiring at least a trigger particle with $p_T>5$ GeV$/c$ in the event, it is interesting to see that TBW4 fit still works well for all these different topological regions. It can be found in the comparison that the particle yield increases with $R_T$ in all topological regions. In the low $R_T$  bin like 0-0.5, the particle yield in the toward and away regions is higher than that in the transverse region, indicating that the event is jet dominant. In high $R_T$ events with $R_T$ around 2.5-5.0, the particle yield in the transverse region is significantly greater than that in the other two regions, suggesting that the jet effect weakens, and softer processes like partonic interactions become more important. In all $R_T$ bins, the $p_T$ spectra in toward and away regions are always harder than the transverse region $p_T$ spectra for different particle species. This difference implies the toward and away regions are more sensitive to the jet productions in the entire event activity range. To further quantitatively analyze the deviation between the fit and the experimental data, we show the pull distribution in Fig.~\ref{fig:pull_region}. From the pull distribution, we observe that the deviations are very small, with all deviations constrained within the three sigma lines, and no significant dependence on the region or $R_T$ is observed. Slightly larger deviations in the pull distribution can be observed in the intermediate $R_T$ regions, especially in the transverse region. 


We perform the same Tsallis Blast-Wave analysis with independent baryon non-extensive parameter to the $p_T$ spectra in high energy proton proton collisions categorized by the unweighted transverse spherocity $S_{O}^{p_{T}=1}$ in three intervals~\cite{ALICE:2023bga}. The TBW4 fits to the pion, kaon and proton $p_T$ data in different spherocity bins and the corresponding pull distributions for the fits are presented in Fig.~\ref{fig:sopt} following the same cosmetics implemented in Fig.~\ref{fig:pt_region} and Fig.~\ref{fig:pull_region}. The results are shown for $S_{O}^{p_{T}=1}$ within 0-100\%, 0-10\%, and 90-100\% categories from left to right, with $p_T$ spectra at the top and the corresponding pull distributions at the bottom. The spherocity classified events are all selected with 0-1\% mid-rapidity hadron yield. High multiplicity events are often driven by MPI effects. To identify the connection between the MPI event structure and the flow like features appearing in high multiplicity events, we utilize the spherocity dependent $p_T$ spectra from events in the top 1\% of the mid-rapidity track number distributions~\cite{ALICE:2023bga}. As focused on the high multiplicity events, the particle yields shown in Fig.~\ref{fig:sopt} are much higher than those shown in Fig.~\ref{fig:pt_region}. Reasonable descriptions for all spherocity events are obtained in this fit. Larger $\chi^2$ is found for the spherocity separated events might suggest that these events contain convoluted effects from multiple physics process and thus have larger fluctuations. More details of the fit parameters can be found in Tab.~\ref{tabTBW4_SOPT}.

In addition, we also compare the fits to the events in various flattenicity bins associated with two different event multiplicity classes~\cite{ALICE:2024vaf}. The results of $p_T$ spectra and the pull distributions for minimum bias events selected with the ALICE V0M detector in 0-100\% percentile are shown in Fig.~\ref{fig:pt_rho} and Fig.~\ref{fig:pull_rho}, respectively. Compared to the case using other event shape selectors, the $p_T$ spectra classified by the flattenicity measurement can be matched to the Tsallis Blast-Wave distribution at a very high precision level. Agreement between data and theoretical fits is found for each particle species along the entire $p_T$ range for all $1-\rho$ bins. Considering that the flattenicity classifier is more sensitive to pile up of the MPI process~\cite{ALICE:2024vaf}, this agreement suggests that the Tsallis distribution is a good approximation to represent the kinematics of a single MPI process. The flattenicity classified high multiplicity events with 0-1\% V0M amplitudes are also included in this analysis, as shown in Fig.~\ref{fig:pt_rhoHM} and Fig.~\ref{fig:pull_rhoHM}. The experimental data align with the model expectations except for some slightly larger deviations in flattenicity classes from 0-1\% to 10-20\% at low $p_T$, indicated by Fig.~\ref{fig:pull_rhoHM}. Table ~\ref{tabTBW4_rho} includes the values for the key parameters obtained in these fits.

	\subsection{Extracted kinetic freeze-out parameters with different event shapes}

	\begin{figure*}[htbp!]
		\begin{center}
			\includegraphics[width=1.0\textwidth]{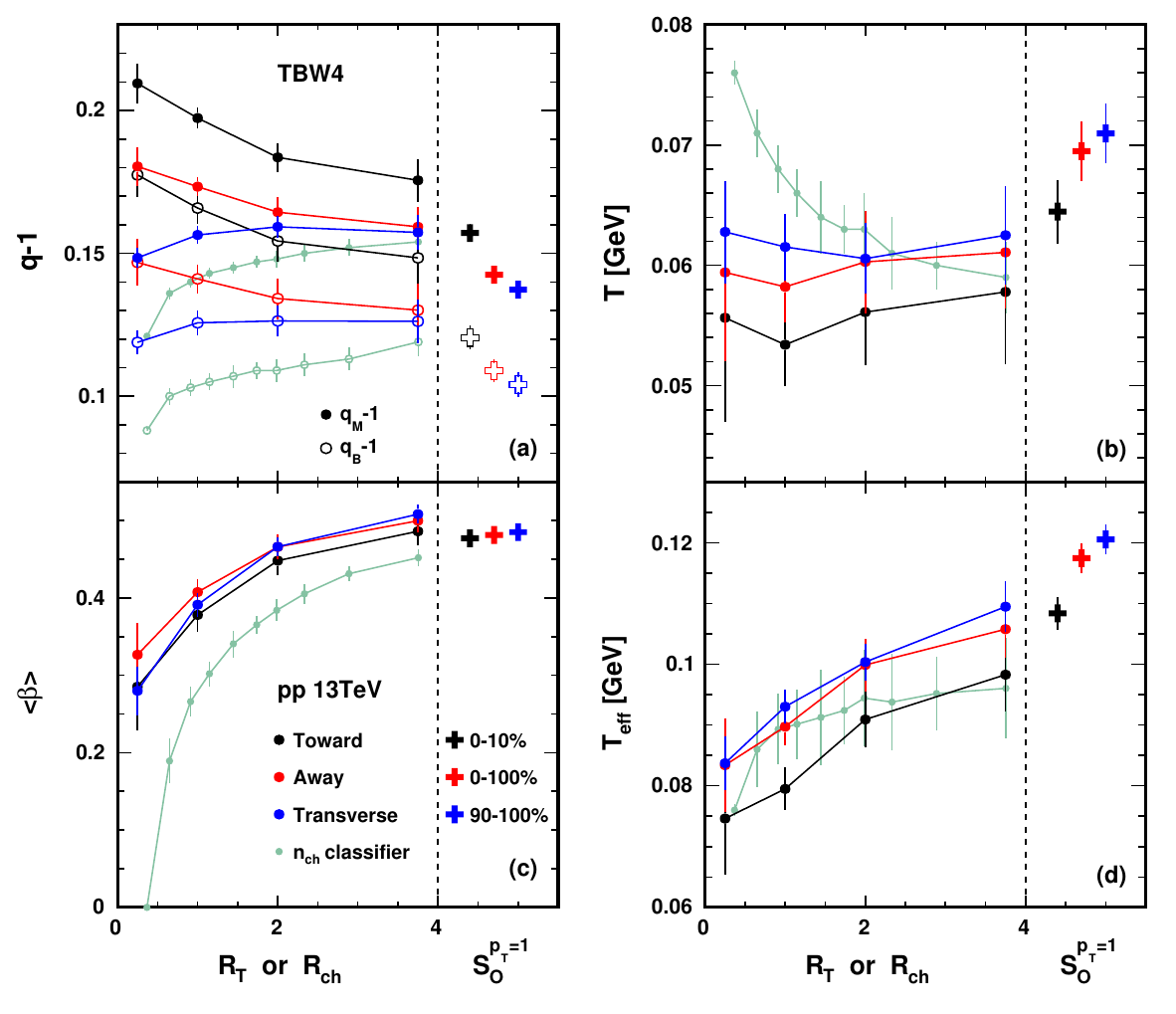}
			     \caption{$R_T$, $R_{ch}$ and $S_O^{{p_T}=1}$ dependence of the extracted freeze-out parameters and the effective temperature $T_{eff}$ in pp collisions at $\sqrt{s}=$ 13 TeV from TBW4 fits. The black, red and blue circle markers correspond to toward, away and transverse region. The green circle markers~\cite{Liu:2022ikt} correspond to $n_{ch}$ classifier. The cross markers correspond to different sphericity. In panel (a), open markers represent the results of $q_B-1$ and solid markers represent the results of $q_M-1$.} 
            \label{fig:rt}
		\end{center}
	\end{figure*}

    \begin{figure*}[htbp!]
		\begin{center}
			\includegraphics[width=1.0\textwidth]{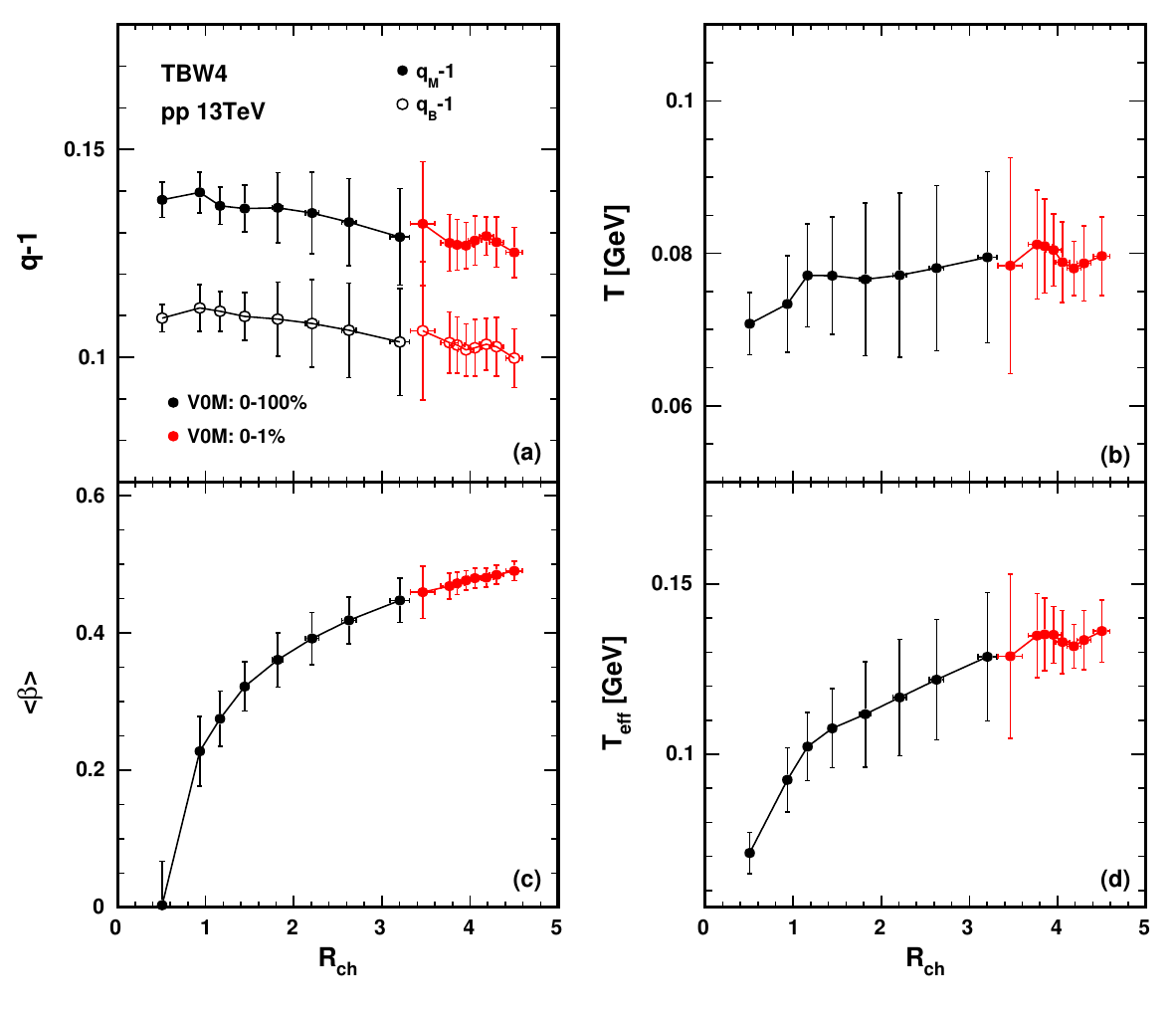}
			     \caption{$R_T$ dependence of the extracted freeze-out parameters and the effective temperature $T_{eff}$ in pp collisions at $\sqrt{s}=$ 13 TeV from TBW4 fits. Black and red represent flattenicity in multiplicity-integrated and high-multiplicity event classes. The direction of increasing $R_T$ is consistent with the direction of increasing flattenicity (from 50-100\% to 0-1\%). In panel (a), open markers represent the results of $q_B-1$ and solid markers represent the results of $q_M-1$.} 
            \label{fig:rt_rho}
		\end{center}
	\end{figure*}
    In this section, we examine the model parameters extracted from the TBW4 fits, including the non-extensive parameter for mesons ($q_M$) and baryons ($q_B$), the average radial flow velocity in the transverse plane ($\langle\beta\rangle$) and the kinetic freeze-out temperature ($T$). The effective temperature \( T_{eff} = \sqrt{\frac{1 + \langle \beta \rangle}{1 - \langle \beta \rangle}} T \) has also been provided to estimate flow and temperature combined effect on $p_T$. If $q$ approaches unity, the Tsallis distribution of the emitting source reduces to an exponential thermal distribution,as typically observed in equilibrium systems. The magnitude of $q-1$ thus serves as a measure to quantify the degree of non-equilibrium effects present in the system. In the following comparisons, in order to confront the trend of extracted parameters from different event shape classifiers on the same basis, we employ the charged particle density information associated with each event shape class and include an event activity like quantity
    \begin{eqnarray}\label{eq:rch}
        R_{ch} = \frac{dN/d\eta}{\langle dN/d\eta \rangle},
    \end{eqnarray}
    where \( dN/d\eta \) represents the charged particle density at a given event category, and \( \langle dN/d\eta \rangle \) denotes the average charged particle density across all the events. In the subsequent calculations involving $R_{ch}$, we use the value of \( \langle dN/d\eta \rangle \) = 6.93$\pm$0.09 from INEL$>$0 events at pp 13 TeV~\cite{ALICE:2020swj} to construct this scaling variable.

    In Fig.~\ref{fig:rt}, to the left of the vertical dashed line, the results for the four parameters obtained with transverse activity classifiers varying with $R_T$ and charged multiplicity classifiers varying with $R_{ch}$ are displayed. The black, red, and blue markers correspond to the results from the toward, away, and transverse regions, respectively. The green markers denote the freeze-out parameters extracted from the inclusive particle $p_T$ spectra classified with event multiplicities obtained from our previous study~\cite{Liu:2022ikt}. It can be seen that the $R_T$ and $R_{ch}$ values for events with different classifiers are within same range 0-4. We also show the parameters obtained with the unweighted transverse spheroicity categorizer in the same figures. As those events are selected with the high multiplicity cut, the $R_{ch}$ is higher than 4 and the difference between the $R_{ch}$ of each spherocity bin is quite small. Therefore, we place the corresponding results at the right of the vertical dashed line in each figure and shift them horizontally as the increasing spherocity. The black, red, and blue cross symbols represent the results from the spherocity intervals 0-10\%, 0-100\%, and 90-100\%, respectively.
    
   The non-extensive parameter values are presented in Fig.~\ref{fig:rt}(a), with solid and open markers representing $q_M$ and $q_B$, respectively. We observe that both $q_M$ and $q_B$ follow a similar $R_T$ or $R_{ch}$ dependence across all topological regions and event shape estimators. Moreover, a clear hierarchy related to event activity can be found, the non-extensive parameter decreases from toward region to transverse region, indicating that non‐equilibrium effects are more pronounced in the jet production zones. The transverse region $q$ increases with $R_T$ similar to the behavior observed for $n_{ch}$ classified events dependent on $R_{ch}$. Conversely, in the toward and away regions, $q$ exhibits a decreasing trend. However, the $q$ values in different topological regions tend to converge at high $R_T$ with the results from $n_{ch}$ classifier at high $R_{ch}$, suggesting that the events become isotropic when multiplicity becomes very high and the MPI effects become very important in jet-dominated regions. The decreasing trend in the toward and away regions reflects a competition between non-equilibrium effects and MPI processes, which evolves with increasing event multiplicity. 

   The freeze-out temperature $T$, as shown in Fig.~\ref{fig:rt}(b), remains nearly constant for all $R_T$ values in each event topological region, despite the substantial uncertainty in the fit parameters. The temperature in toward region is systematically smaller than that in away and transverse regions. Unlike the $n_{ch}$ classifier, where $T$ decreases with $R_{ch}$, the temperature extracted by the transverse activity selector is generally lower. This difference is likely due to the $p_T>5$ GeV/$c$ requirement imposed when selecting events based on transverse activity. It is thus not surprising to see that the radial flow velocity, sensitive to the transverse energy density, is significantly higher when using transverse activity selector compared to the $n_{ch}$ estimator, as shown in Fig.~\ref{fig:rt}(c). The radial flow velocity from different topological regions is close to each other and increases with $R_T$, starting from a non-zero value even at low $R_T$. The effective temperature presented in Fig.~\ref{fig:rt}(d), which encapsulates both thermal motion and collective transverse expansion, is observed to increase with rising transverse activity. This trend is primarily attributed to the rapid growth of radial flow velocity with increasing $R_T$, indicating that MPI effects play a significant role in shaping the transverse momentum spectra across all event topologies.  
   
   On the other hand, applying the spherocity classifier reveals a systematic decline in the Tsallis non-extensivity parameter as one moves from pencil-like events (0–10\% spherocity percentile) to sphere-like events (90–100\% percentile). Notably, the $q$ values extracted for sphere-like selections are substantially lower than those obtained via alternative classification methods, underscoring spherocity’s ability to isolate near-equilibrium, isotropic particle distributions. However, the kinetic freeze-out temperature is found to be slightly larger in more isotropic events than in jetty events. This trend mirrors observations from transverse activity studies, where jet-associated regions show higher $q$ and lower temperatures. 
   The radial flow velocity appears to be insensitive to the spherocity selection and reaches saturation, as these events already correspond to high–multiplicity collisions with very large transverse energy density. The effective temperature turns out to be increasing from jetty events to isotropic events because of the enhancement of the kinetic freeze-out temperature rather than changes in the flow velocity.

    The same kinematic freeze-out parameter distributions obtained for the flattenicity estimators are presented in Fig.~\ref{fig:rt_rho}. The results for the multiplicity integrated events and 0-1\% V0M high multiplicity events are shown with black and red markers, respectively. The results are presented as a function of $R_{ch}$. As $R_{ch}$ increases, flattenicity grows likewise, indicating events shift to more uniform distributions. Figure ~\ref{fig:rt_rho}(a) shows the non-extensive parameters for meson ($q_M$) and baryon ($q_B$) in solid and open markers. The non-extensive parameters remain nearly constant across different flattenicity bins within uncertainties. Additionally, the values of $q_M$ and $q_B$ closely match those observed in spherocity identified events with the highest isotropy within 90-100\% $S_O^{p_T=1}$ percentile. The freeze-out temperature shown in Fig.~\ref{fig:rt_rho}(b) is found to be also flat and close to the value obtained in isotropic events. This value is generally higher than the temperature extracted from the high multiplicity $n_{ch}$ selected events. As the flattenicity estimator is expected to be less biased toward the high $p_T$ jet effects and focusing on the MPI related soft QCD dynamics, this constancy implies that the hadron productions in pp collisions decouples at similar local statistical freeze-out conditions, regardless of the event multiplicity. The radial flow velocity shown in Fig.~\ref{fig:rt_rho}(c) increases with $R_{ch}$ (and with flattenicity) similar to the observation found in general multiplicity or transverse activity estimators. The effective temperature displayed in Fig.~\ref{fig:rt_rho}(d) rises with $R_{ch}$ in the multiplicity integrated events, driven by enhanced flow velocity. In high multiplicity events (0-1\% V0M selection), the effective temperature remains relatively stable, reflecting minimal variations in $p_T$ spectra with respect to flattenicity~\cite{ALICE:2024vaf}. Overall, the flattenicity classifier effectively decouples jet-related hadronization effects reflected in the non-extensive parameter and local temperature from collective flow dynamics driven by stacked MPI processes, providing a clean probe of the soft QCD freeze-out stage.

Determining the onset of collectivity in small systems is crucial for understanding if they can achieve the critical energy density necessary for QGP formation. This type of investigation is usually carried out using global multiplicity to estimate the energy density of the system. A prominent approach to this search, exemplified by the work of Ref.~\cite{Olimov:2021ogj,Olimov:2021tit}, employed Tsallis and Hagedorn functions to fit the hadron spectrum in pp collisions, identifying an inflection point in the transverse flow velocity as a possible phase transition signal. Their analysis suggests a smooth evolution toward thermalization at high event activity similar to our findings in this work.
It is interesting to note that the current work by incorporating event shape observables offers a more differential probe of the medium’s properties than multiplicity alone. We argue that the emergence of collective effects is not solely dependent on the global energy density probed by multiplicity, but also on its spatial distribution within the collision volume. Event shape classifiers enable comparisons between events with similar multiplicities but vastly different internal geometries, providing complementary dimension to works in Ref.~\cite{Olimov:2021ogj,Olimov:2021tit}.

    \subsection{Parameter correlation}
	
	\begin{figure*}[htbp!]
		\begin{center}
			\includegraphics[width=1.0\textwidth]{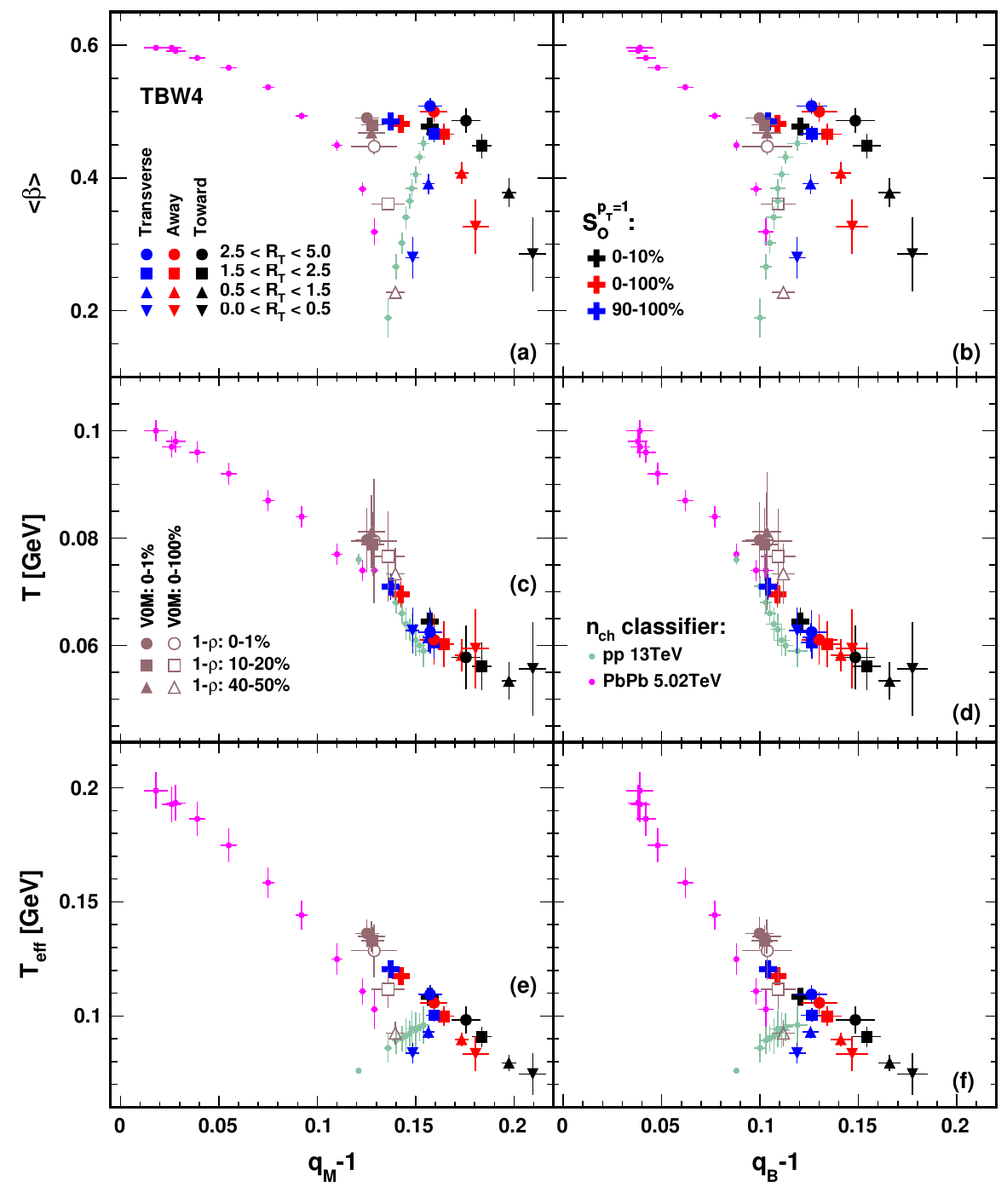}
			\caption{$\langle\beta\rangle$ vs $q-1$, $T$ vs $q-1$ and $T_{eff}$ vs $q-1$ in pp collisions at $\sqrt{s}=$ 13 TeV from TBW4 fits. $q$ includes $q_M$ (left column) and $q_B$ (right column). The results for transverse event activity, spherocity, flattenicity and $n_{ch}$ classifier have been shown, with different colors and shapes denoting the various event classes. The $n_{ch}$ classifier results are taken from our previous work~\cite{Liu:2022ikt}.
			} \label{fig:pars}
		\end{center}
	\end{figure*}

    In the context of non-equilibrium statistics, temperature and flow velocity can be associated with viscosity through linear or quadratic dependencies on the non-extensive parameter. Figure~\ref{fig:pars} displays the results of $\langle\beta\rangle$ and $T$ varying with $q-1$ from the TBW4 fits with different event shape estimators in pp collisions at $\sqrt{s}=13$ TeV. The transverse event activity dependences in toward, away and transverse region are shown in different colors with the marker styles representing each $R_T$ bin. The spherocity identified results are demonstrated by the crossing markers with colors showing different spherocity bins. The flattenicity related results are made in brown color with open and solid markers indicating minbias events and 0-1\% V0M high multiplicity events. Meanwhile, we plot the $n_{ch}$ classified results for pp collisions at $\sqrt{s}=13$ TeV and PbPb collisions at $\sqrt{s_{NN}}=5.02$ TeV from our previous work~\cite{Liu:2022ikt} in the same figure with cyan and magenta symbols to compare the system size effect. The first row shows the radial flow velocity correlation with the non-extensive parameter. It is seen in this figure that the different event topological region results converge from low $R_T$ to high $R_T$ events, implying the high $R_T$ events from all topological regions are becoming dominanted by the same underlying event effects. However, the results for the transverse region evolve from low $q$ to high $q$ as the flow velocity grows with $R_T$, suggesting the transverse region receives strong fluctuation effects in high activity events, similar to the observations in $n_{ch}$ classified pp collision results. The same enhancement of hard process bias effects in high multiplicity events may apply to both scenarios to account for this similarity. The $\langle\beta\rangle$ versus $q$ correlation in toward and away region shifts along the opposite direction, leading to smaller $q$ when $\langle\beta\rangle$ increases, which can be understood as an outcome of the interplay between the hard process and the underlying event soft physics varying with the event activity. 
    The spherocity classified events are restricted to the high multiplicity events, so $\langle\beta\rangle$ remains almost unchanged at a velocity as high as that in the high $R_T$ events while $q_M$ and $q_B$ reduces from azimuthally jetty events (0-10\%) to more isotropic events (90-100\%) approaching equilibrium. The flattenicity categorized events are found to be only changing in $\langle\beta\rangle$ with the non-extensive parameters being stable when the flatness of the system increases. It is interesting to note that the behavior of the flattenicity driven evolution effects in pp collisions aligns with the large system PbPb collision results.  
    
    The kinetic freeze-out temperature correlation with the non-extensive parameter has been shown in the second row. A significant universal scaling feature has been observed across different collision systems with various event shape categorizers. This universal evolution curve in the $T$ vs $q-1$ plane reinstates our previous finding of the existence for the universality in the kinetic freeze-out features independent of the collision system, implying that a unified partonic evolution stage, where similar QCD dynamics govern parton interactions and evolution, leads to consistent freeze-out parameters. The $1-\rho$ identified events are found to have the largest $T$ with very small variations between different flattenicities. The jet induced effects amplified by the toward and away event topological selections in different event activity are delivering very small $T$ and large $q$ representing strong dynamical fluctuations in temperature due to hard process effects. The low $R_T$ toward region approaches very small $T$ with large $q$. 

    In the third row, we present the effective temperature versus $q$ correlations. A universal upper boundary determined by the maximum radial flow velocity with variations related to the temperature can be found from central PbPb collisions with the largest $T_{eff}$ and smallest $q$ to the jetty toward region results with the smallest $T_{eff}$ and largest $q$. The diverged radial flow velocity dependence in each event shape selection leads to evolutions along different branches under the common boundary due to the energy density constrained by the system size.

    It has been observed in this comparison that the statistical freeze-out parameter space has been largely expanded with different event shape measurements. Event shapes can tag the different limits of the freeze-out parameter space for pp collisions. It is also shown that there is a maximum kinetic freeze-out temperature and flow velocity which can be reached in pp collisions with all different event shapes at a level close to the peripheral PbPb collisions. This difference represents the enhancement of inter-nucleon dynamics in generating the initial energy and entropy densities compared to the sub-nucleon partonic fluctuations induced by color glass condensate effects. In all these different event shape observables, the flattenicity shows unique features of mimicking the collective motion effects observed in PbPb collisions by overlapping multiple parton interactions in an additive way, making this observable of special interest to isolate the soft interactions dominated flow physics in small systems.

	\section{Summary}
	\label{sec:summary}

    In this study, we employed the TBW4 model with the independent baryon non-extensive parameter $q_B$, to fit the $p_T$ spectra of $\pi$, $K$, and $p$ in $\sqrt{s} = 13$ TeV pp collisions with different event shapes. Utilizing event shape classifiers including transverse event activity, unweighted transverse spherocity, and flattenicity, we explored the connection between event shape and particle emission characteristics in pp collisions. Across all classifiers and topological regions, TBW4 describes the spectra relatively well, even when selecting events with a high $p_T$ leading track in event activity studies.

The non‐extensive parameters decrease with increasing isotropy or MPI activity, indicating that more isotropic, MPI-dominated events approach equilibrium.
A stronger fluctuation effect due to the nonequilibrium dynamics represented by the magnitude of the nonextensive parameter is found in the toward and away regions compared to the transverse region results. The radial flow velocity increases with the activity or multiplicity of the event in all different topological regions of the event with various event shapes and saturates in high‐multiplicity bins, reflecting stronger collective expansion in denser underlying events.
Freeze-out parameter values from different event topological regions relative to the leading particle converge at high activity, indicating a common soft-physics limit across topologies.
While the kinetic freeze-out temperature $T$ shows less variation across event classes, a universal scaling in the $T$ vs $q-1$ plane is observed, consistent with previous findings and demonstrating a common freeze‐out physics governed by entropy production and multi‐parton interactions~\cite{Liu:2022ikt}.
Comparisons with PbPb collisions highlight similarities in the maximum achievable collective effects in pp events and peripheral heavy-ion collisions. 
Notably, the flattenicity classifier effectively decouples jet-related hadronization effects and local temperature from collective flow dynamics, providing a cleaner probe of the soft QCD freeze-out stage.
The non-extensive parameters and freeze-out temperature remain nearly constant across different flattenicity bins, similar to values in highly isotropic spherocity identified events.

    The study explores the interplay of hard partonic scattering and soft processes during the development of flow-like signatures. 
These findings highlight the effectiveness of event shape classifiers in probing soft QCD processes and collective phenomena in small systems, providing valuable insights into the emergence of collective phenomena and the nature of the freeze-out stage in high-energy proton-proton collisions.
 	
	\begin{acknowledgments}
		We would like to thank Zebo Tang, Wangmei Zha and Qiye Shou for helpful discussions. This work was supported by the National Key Research and Development Program of China (Grant No. 2024YFA1610800), the National Natural Science Foundation of China (Nos. 12205259, 12147101, 12275103, 12061141008), the Fundamental Research Funds for the Central Universities, China University of Geosciences(Wuhan) with No. G1323523064 and the Innovation Fund of Key Laboratory of Quark and Lepton Physics QLPL2025P01.
		
	\end{acknowledgments}
	
	
	\bibliography{main}
	
	\clearpage
	
	\onecolumngrid
	
	\section*{Appendix}
        \subsection{Fit parameters}
	
	\begin{table*}[htbp]
		\caption{\label{tabTBW4} Extracted kinetic freeze-out parameters and $\chi^{2}/nDoF$ from TBW4 fits to identified particle transverse spectra in pp collisions at $\sqrt{s}=$ 13 TeV of different region and $R_T$.}
		\begin{ruledtabular}
			\resizebox{.97\columnwidth}{!}{
				\begin{tabular}{ccccccc}
					$\rm Region$  & $R_{T}$ & $\langle\beta\rangle$  &  $T \;(\rm{MeV})$  & $ q_{M}-1$ & $q_{B}-1$ & $\chi^{2}/nDoF$ \\
					\hline
                    Toward & 0.0$<R_{T}<$0.5	& 0.285$\pm$0.056 & 56$\pm$9 & 0.209$\pm$0.007 & 0.177$\pm$0.008 & 7/68 \\
                           & 0.5$<R_{T}<$1.5	& 0.378$\pm$0.022 &	53$\pm$3 & 0.197$\pm$0.004 & 0.166$\pm$0.006 & 19/68 \\
                           & 1.5$<R_{T}<$2.5	& 0.448$\pm$0.018 &	56$\pm$4 & 0.184$\pm$0.005 & 0.154$\pm$0.007 & 16/68 \\
                           & 2.5$<R_{T}<$5.0	& 0.486$\pm$0.019 &	58$\pm$6 & 0.176$\pm$0.008 & 0.148$\pm$0.010 & 11/68 \\
                    \hline
                    Away & 0.0$<R_{T}<$0.5 & 0.327$\pm$0.041	& 59$\pm$7 & 0.180$\pm$0.007 & 0.147$\pm$0.008 & 5/68 \\
                         & 0.5$<R_{T}<$1.5 &	0.408$\pm$0.017	& 58$\pm$3 & 0.173$\pm$0.003 & 0.141$\pm$0.005 & 24/68 \\
                         & 1.5$<R_{T}<$2.5 &	0.466$\pm$0.016	& 60$\pm$4 & 0.164$\pm$0.005 & 0.134$\pm$0.007 & 15/68 \\
                         & 2.5$<R_{T}<$5.0 &	0.500$\pm$0.015	& 61$\pm$5 & 0.159$\pm$0.007 & 0.130$\pm$0.009 & 11/68 \\
                    \hline
                    Transverse & 0.0$<R_{T}<$0.5 & 0.280$\pm$0.031 & 63$\pm$4 & 0.148$\pm$0.003 & 0.119$\pm$0.004 & 15/68 \\
                               & 0.5$<R_{T}<$1.5	& 0.391$\pm$0.015 & 62$\pm$3 & 0.156$\pm$0.003 & 0.126$\pm$0.004 & 29/68 \\
                               & 1.5$<R_{T}<$2.5	& 0.466$\pm$0.012 &	61$\pm$3 & 0.159$\pm$0.004 & 0.126$\pm$0.006 & 31/68 \\
                               & 2.5$<R_{T}<$5.0	& 0.508$\pm$0.012 &	63$\pm$4 & 0.157$\pm$0.006 & 0.126$\pm$0.008 & 17/68 \\
				\end{tabular}
			}
		\end{ruledtabular}
	\end{table*}

	\begin{table*}[htbp]
		\caption{\label{tabTBW4_SOPT} Extracted kinetic freeze-out parameters and $\chi^{2}/nDoF$ from TBW4 fits to identified particle transverse spectra in pp collisions at $\sqrt{s}=$ 13 TeV of different $S_{O}^{p_{T}=1}$.}
		\begin{ruledtabular}
			\resizebox{.97\columnwidth}{!}{
				\begin{tabular}{cccccc}
					
                    $S_{O}^{p_{T}=1}$  & $\langle\beta\rangle$  &  $T \;(\rm{MeV})$  & $ q_{M}-1$ & $q_{B}-1$ & $\chi^{2}/nDoF$ \\
					\hline
                    $0-100\%$ & 0.481$\pm$0.009 & 70$\pm$2 & 0.143$\pm$0.003 & 0.109$\pm$0.004 & 43/77 \\
                    $0-10\%$ & 0.477$\pm$0.009	& 64$\pm$3 & 0.157$\pm$0.003 & 0.120$\pm$0.004 & 42/77 \\
                    $90-100\%$ & 0.485$\pm$0.008 & 71$\pm$2 & 0.137$\pm$0.003 & 0.104$\pm$0.004 & 46/77 \\
			 \end{tabular}
			}
		\end{ruledtabular}
	\end{table*}

	\begin{table*}[htbp]
		\caption{\label{tabTBW4_rho} Extracted kinetic freeze-out parameters and $\chi^{2}/nDoF$ from TBW4 fits to identified particle transverse spectra in pp collisions at $\sqrt{s}=$ 13 TeV of different V0M and $1-\rho$.}
		\begin{ruledtabular}
			\resizebox{.97\columnwidth}{!}{
				\begin{tabular}{ccccccc}
					$\rm V0M$  & $1-\rho$ & $\langle\beta\rangle$  &  $T \;(\rm{MeV})$  & $ q_{M}-1$ & $q_{B}-1$ & $\chi^{2}/nDoF$ \\
					\hline
                    $0-100\%$ & $0-1\%$     & 0.447$\pm$0.032 & 79$\pm$11 & 0.129$\pm$0.012 & 0.104$\pm$0.013 & 3/77 \\
                            & $1-5\%$     & 0.418$\pm$0.034 & 78$\pm$11 & 0.133$\pm$0.011 & 0.106$\pm$0.011 & 3/77 \\
                            & $5-10\%$    & 0.392$\pm$0.038 & 77$\pm$11 & 0.135$\pm$0.010 & 0.108$\pm$0.011 & 3/77 \\
                            & $10-20\%$   & 0.361$\pm$0.040 & 77$\pm$10 & 0.136$\pm$0.008 & 0.109$\pm$0.009 & 3/77 \\
                            & $20-30\%$   & 0.322$\pm$0.036 & 77$\pm$8 & 0.136$\pm$0.006 & 0.110$\pm$0.006 & 5/77 \\
                            & $30-40\%$   & 0.275$\pm$0.040 & 77$\pm$7 & 0.136$\pm$0.004 & 0.111$\pm$0.005 & 6/77 \\
                            & $40-50\%$   & 0.228$\pm$0.051 & 73$\pm$6 & 0.140$\pm$0.005 & 0.112$\pm$0.006 & 6/77 \\
                            & $50-100\%$  & 0.003$\pm$0.064 & 71$\pm$4 & 0.138$\pm$0.004 & 0.109$\pm$0.003 & 6/77 \\
                    \hline
                    $0-1\%$   & $0-1\%$     & 0.490$\pm$0.014 & 80$\pm$5 & 0.125$\pm$0.006 & 0.100$\pm$0.007 & 11/77 \\
                            & $1-5\%$     & 0.484$\pm$0.014 & 79$\pm$5 & 0.128$\pm$0.006 & 0.103$\pm$0.007 & 11/77 \\
                            & $5-10\%$    & 0.481$\pm$0.014 & 78$\pm$4 & 0.129$\pm$0.005 & 0.103$\pm$0.006 & 11/77 \\
                            & $10-20\%$   & 0.480$\pm$0.014 & 79$\pm$5 & 0.128$\pm$0.006 & 0.102$\pm$0.007 & 10/77 \\
                            & $20-30\%$   & 0.476$\pm$0.014 & 80$\pm$5 & 0.127$\pm$0.006 & 0.102$\pm$0.006 & 8/77 \\
                            & $30-40\%$   & 0.472$\pm$0.016 & 81$\pm$6 & 0.127$\pm$0.006 & 0.103$\pm$0.007 & 6/77 \\
                            & $40-50\%$   & 0.468$\pm$0.019 & 81$\pm$7 & 0.128$\pm$0.007 & 0.104$\pm$0.007 & 5/77 \\
                            & $50-100\%$  & 0.459$\pm$0.038 & 78$\pm$14 & 0.132$\pm$0.015 & 0.106$\pm$0.017 & 2/77 \\
				\end{tabular}
			}
		\end{ruledtabular}
	\end{table*}

\end{document}